# Wireless Mobility Management with Prediction, Delay Reduction and Resource Management in 802.11 Networks

Biju Issac, *Member, IAENG*, Khairuddin Hamid, and C. E. Tan

*Abstract*—In 802.11 wireless infrastructure networks, as the mobile node moves from one access point to another, the active connections will not be badly dropped if the handoff is smooth and if there are sufficient resources reserved in the target access point. In a 5x5 grid of access points, within a 6x6 grid of regions, by location tracking and data mining, we predict the mobility pattern of mobile node with good accuracy. The pre-scanning of mobile nodes, along with pre-authenticating neighbouring access points and pre-reassociation is used to reduce the scan delay, authentication delay and re-association delay respectively. The model implements first stage reservation by using prediction results and does second stage reservation based on the packet content type, so that sufficient resources can be reserved when the mobile node does the handoff to the next access point. The overall mobility management scheme thus reduces the handoff delay. The performance simulations are done to verify the proposed model.

## I. INTRODUCTION

As the mobile node moves across access points in an 802.11 network, the mobility management becomes a challenging area of research as the mobile node should enjoy smooth handoff and low packet drop. Mobility management covers the options for storing and updating the location information of mobile users who are connected to the system and allocating the right amount of resources. Once an accurate mobile path prediction can be done, the predicted path or movement can help increase the efficiency of wireless network, by effectively allocating resources to the most probable access point (that can be the next point of attachment) instead of blindly allocating excessive resources in the mobile user's neighborhood [1], [2]. In this paper we want to come up with a predictive mobility management system which could eventually make mobility on an 802.11 network more proactive with minimum loss and delay, when compared to existing schemes.

This paper is organized as follows. Section 2 is the related works on mobility management, section 3 is the proposed mobility management architecture, section 4 is the predictive mobility management blocks, section 5 is the overall predictive mobility management algorithm, section 6 is the performance simulation results for mobility prediction, section 7 is the related works on delay and resource management, section 8 is the delay and resource management, section 9 is performance simulation results for delay and resource management and section 10 is the conclusion.

## II. RELATED WORKS ON MOBILITY MANAGEMENT

Park et al. has proposed an Adaptive Handover Control Architecture (AHCA) in [3]. As a core part of the architecture proposed by Park, the *Adaptive Handover Engine* takes inputs from several input pre-processing modules (like network resource information from the *Network Resource Prober*, traffic QoS (Quality of Service) attributes from the *Traffic Classifier*, user preferences information from the *User Input Handler*, and policy information from the *Policy Input Handler)*. Then, it selects the best handover mechanism using a handover adaptation algorithm, so that the chosen handover strategy produces the best performance for the user. The architecture has been inspired by the related research in the field of handoff control, such as Programmable Handoffs [4], Policy-Enabled Handoffs [5] and other adaptive or feedback-based control approaches [6].

Liu and Maguire have proposed a predictive mobility management algorithm for supporting mobile computing and communications [7]. This predictive algorithm deploys the regular-pattern detection algorithms to decompose the complicated daily movement into two parts: regular pattern part and random movement part. The Movement Circle (MC) and Movement Track (MT) model is proposed and used to predict the regular movement, while the random parts of the movement is modeled and predicted by the Markov chain model. Sabeur et al. proposed a Mobile Party protocol, which is a new scheme for mobility management in the context of mesh networks, where it integrates routing and mobility management functionalities together [8]. They claim that it provides a

Manuscript received July 24, 2008.
B. Issac is with the School of Computing and Design, Swinburne University of Technology (Sarawak Campus), Kuching, Sarawak, Malaysia (phone: +60-82-416353; fax: +60-82-423594; e-mail: bissac@swinburne.edu.my).
K. A. Hamid is with University Malaysia Sarawak, 94300 Kota Samarahan, Sarawak, Malaysia (e-mail: khair@cans.unimas.my).
C. E Tan is with the Faculty of Computer Science and Information Technology, University Malaysia Sarawak, 94300 Kota Samarahan, Sarawak, Malaysia (e-mail: cetan@fit.unimas.my).

viable solution for mobility and routing management, assuring scalability and seamless communications.

Mihovska et al. investigated [9] the policy-based mobility management scheme proposed for the WINNER system (a resource management system). The paper describes the cooperation architecture and mobility management functionalities and the rules that are applied for the support of mobility within the above scenarios. The rules are applied to solve problems such as: what is the highest and lowest priority traffic, what are the levels in between, and how are they differentiated, how do we guarantee delivery of highest priority traffic (e.g., real-time applications), and how can we guarantee the required for the delivery bandwidth. Location determination can also be scaled into different levels by using different positioning methods. Further, the paper describes other actions executed by the implemented policy-based mobility management scheme like radio access technology (RAT) association, user and flow context transfer, handover decision, and deployment priority. Uskela gives a framework that shows how the different definitions of the mobility management relate to each other. Further, it discusses on how some existing mobility management solutions map into the new framework [10].

Sun and Sauvola discusses the effects of mobility on both the architectures and protocols for network communications [11]. Mobility management is defined as two complementary operations, i.e. location management and handoff management. The processing stages of the two operations are introduced, together with the analyses of key research issues and possible solutions. Finally, the paper discusses issues concerning the performance evaluation of mobility management schemes.

### III. PROPOSED MOBILITY MANAGEMENT ARCHITECTURE

We would like to propose a micro mobility handoff mechanism that is fast and adaptive, which we call as Predictive Mobility Management Scheme (PMMS). The proposal in [3] was for IPv4 networks and did not use mobility prediction and pre-authentication. Our focus is more on WLAN (802.11) installations within a restricted campus. The main objective of the proposal is to predict the mobility path of a mobile node and use that information to lessen the handoff delay. Various other delays encountered during handoff – like probe delay, authentication delay and reassociation delay would be handled under delay management. Even resource management is done during handoff using two stages of resource reservation. These concepts would be explained as we progress further.

Consider the access points numbered 0 to 24, arranged in a 5 x 5 matrix form. Surrounding each access point (AP) there are four regions. These regions are numbered from R0 to R35 and is positioned around the 25 access points, as a 6 x 6 matrix as shown in the fig. 1.

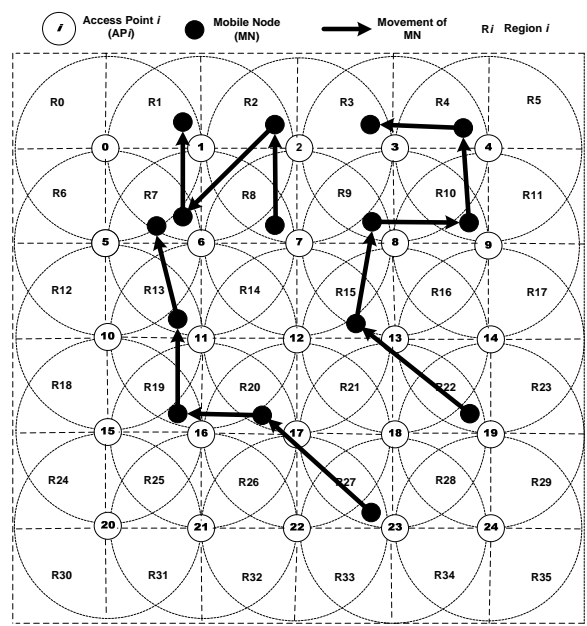

Figure 1. Access Points and Mobile regions in a matrix format, showing three mobile node paths

The mobile node (MN) can move around any AP in the surrounding four regions or can move to other adjacent regions of neighbouring AP. A Mobile Path Prediction Server (MPPS) is connected to the Local Area Network where the access points are connected, to make the mobility prediction in a centralized way. At frequent intervals of time, the mobile node transmits the Received Signal Strength Indication (RSSI) of the surrounding APs to the MPPS.

### IV. PREDICTIVE MOBILITY MANAGEMENT BLOCKS

The functional blocks shown in fig. 2 can be identified to form a mobility management system with smooth, fast and adaptive handoff as the mobile node moves from one access point (AP) to another. The Predictive Mobility Management Scheme (PMMS) architecture as shown in fig. 2 has the following blocks – Mobility Prediction Block, Delay Management Block with sub-blocks like Authentication Delay Management Block, Scan Delay Management Block and Reassociation Delay Management Block, Resource Reservation Block and Handoff block.

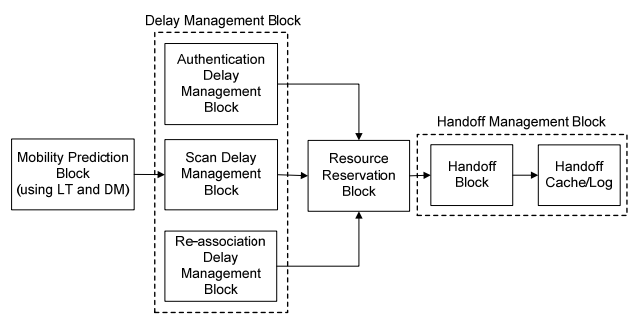

Figure 2. The PMMS architecture

### A. Mobility Prediction Block

We propose a centralized mobility prediction technique that is a hybrid version of location tracking without GPS and data mining technique called *Location Tracking with Data Mining Prediction Scheme* (LTDMPS). Location tracking is done through a central server that receives the RSSI (Received Signal Strength Indication) data regarding neighbouring APs from the mobile node which does regular active scanning.

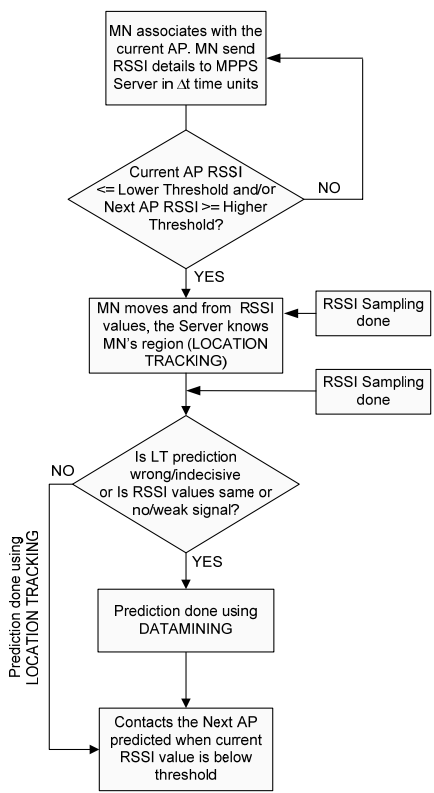

Figure 3. General Mobility Prediction Block

Data mining is done by the same server by using the mobility path history of mobile node movements and by extracting mobility patterns from that [12]. In location tracking, by seeing the details of communication of AP's identity and RSSI value, the MPPS server can know which region the mobile node is located. All valid mobile paths by different users or mobile nodes are recorded (along with regions) in a central database in MPPS server.

We assume that the MN moves from one region to the other and pauses for a duration of *t* time units in any region, before making any movements further. By seeing the details of communication of AP's identity and RSSI value, the MPPS server can know which region the mobile node is located. All valid mobile paths by different users or mobile nodes are recorded (along with regions) in a central database in MPPS server. There can be some indecisive or failure scenarios as follows. The RSSI values received can be very weak or corrupted from all the neighbouring APs or the RSSI values received are similar in value and no clear prediction can be made on the next AP. As RSSI sampling is done frequently, the MN or server can know whether the location tracking prediction result is true or false. If the prediction is false (as the MN moves to other AP) or if there is an indecisive scenario, it moves to data mining mode. The Data mining approach consists of three phases: user mobility pattern mining, generation of mobility rules using the mined user mobility patterns, and the mobility prediction. The next inter-cell movement of mobile users is predicted based on the mobility rules in the last phase. The flow chart of the mobile prediction block is shown in fig. 3.

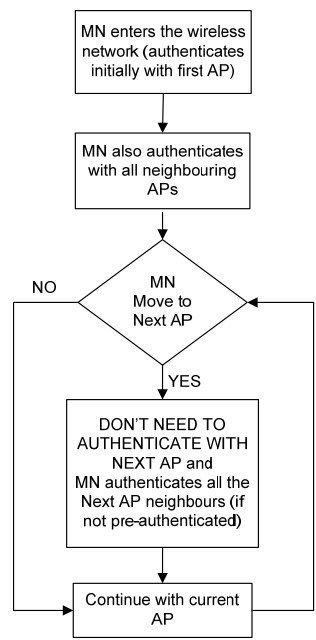

Figure 4. Authentication Delay Management Block

### B. Delay Management Block

This block contains three sub-blocks namely – Authentication Delay Management Block, Scan Delay Management Block and Reassociation Delay Management Block. An empirical measurement of these delays can be found in the paper by Mishra et al [13].

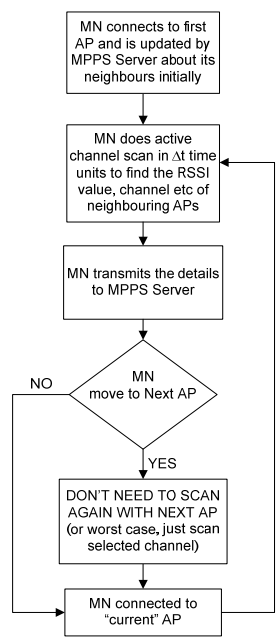

Figure 5. Probe/Scan Delay Management Block

Under authentication delay management, pre-authentication with all the neighboring access points (of the current access point) would happen when mobile node gets connected to a current access point. So when the mobile node moves to the next AP, there is no need to further authenticate and hence can cut down on authentication delay [14]. The flow chart of the authentication delay management block is shown in fig. 4.

This block segregates traffic based on ToS (Type of Service) field in IPv4 (or the Traffic Class and Flow Label fields in IPv6 networks). In case of IPv6 packets, the traffic class field facilitates handling of real-time data and any other data that requires special handling. This field can be used by mobile nodes and forwarding routers (like Access Routers) to identify and distinguish between different classes or priorities of IP packets. Flow label field distinguishes packets that require the same treatment, in order to facilitate the handling of real-time traffic.

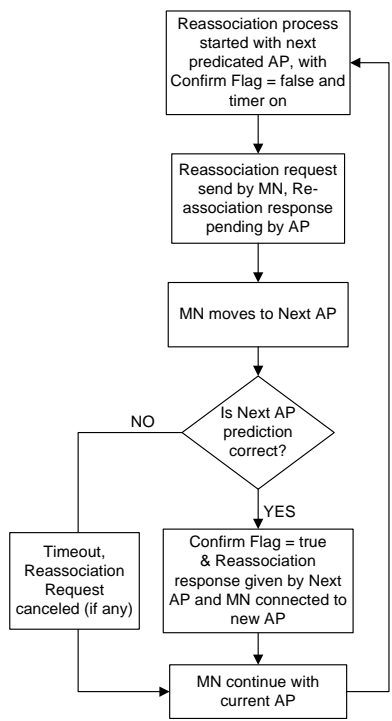

Figure 6. Reassociation Delay Management Block

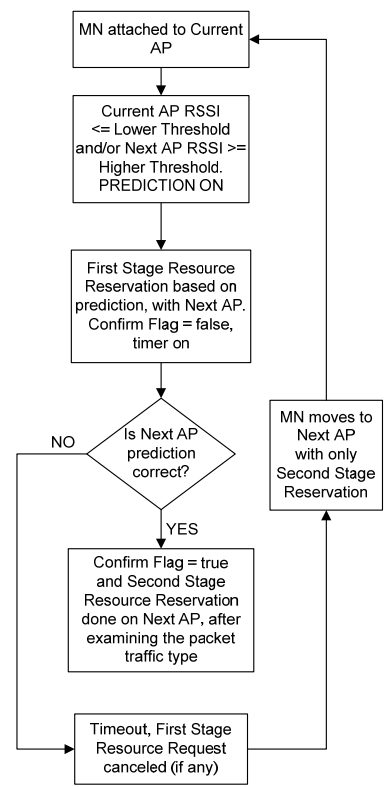

Figure 7. Resource Reservation Block

Under probe or scan delay management, pre-scanning by the MN happens in Δt time units. During handoff, the MN already knows the details of the next AP and the time consuming scan during handoff is avoided [15] - [17]. The flow chart of the probe delay management block is shown in fig. 5.

Pre-reassociation is done (half cycle), where by the reassociation request is sent based on the mobility prediction. Only when next AP prediction is confirmed true and MN confirms it, that the reassociation process is completed, by sending the reassociation response. Some works done by researchers are along similar lines, though not exactly to the same grid scenario as in the paper by Ping-Jung et al. [18]. The flow chart of the reassociation delay management block is shown in fig. 6.

### C. Resource Reservation Block

Here two stages of reservations are done on the next access point that the mobile node is planning to connect. First-Stage Reservation (with timer on) is done based on the mobility prediction. Second Stage reservation is done based on the packet traffic type (text or audio), once the prediction is confirmed, just before handoff.

The reservation made can be categorized into 2 groups –*active* when MN is using it currently [19-[20] and *passive* when the resources are reserved, but not used. To increase the resource utilization we need to allow other hosts to use the passive reservation temporarily and to preempt them once the MN needs it [19, 21]. This can help utilize the resources better which would otherwise be wasted from pre-reservation. The flow chart of resource reservation is shown in fig. 7.

### D. Handoff Management Block

After issuing the RSSI handoff threshold warning, the actual handoff takes place. If the resources reserved are sufficient for the accessing MN, then the MN enjoys a smooth handoff. As an option, if extra resources need to be reserved, it can be released from an emergency store. If the reservation of resources is not a success, then the MN (that is accessing or sending audio data) may lose packets if it moves from the current AP.

A Handoff cache/log can be constructed to store the previous handoff decisions, to speed up future decisions. Some static attributes from the past handoff decisions can be reused, instead of finding it again thus reducing latency during handoff. This cache or log would be referred to as past history of the mobile node movements.

## V. OVERALL PREDICTIVE MOBILITY MANAGEMENT ALGORITHM

The flow chart in fig. 8 shows the algorithm of the Predictive Mobility Management Scheme (PMMS) with optimized delays and resource reservation. It has been shown in three stages – Pre-mobile stage, Mobile Stage and Post-Mobile Stage.

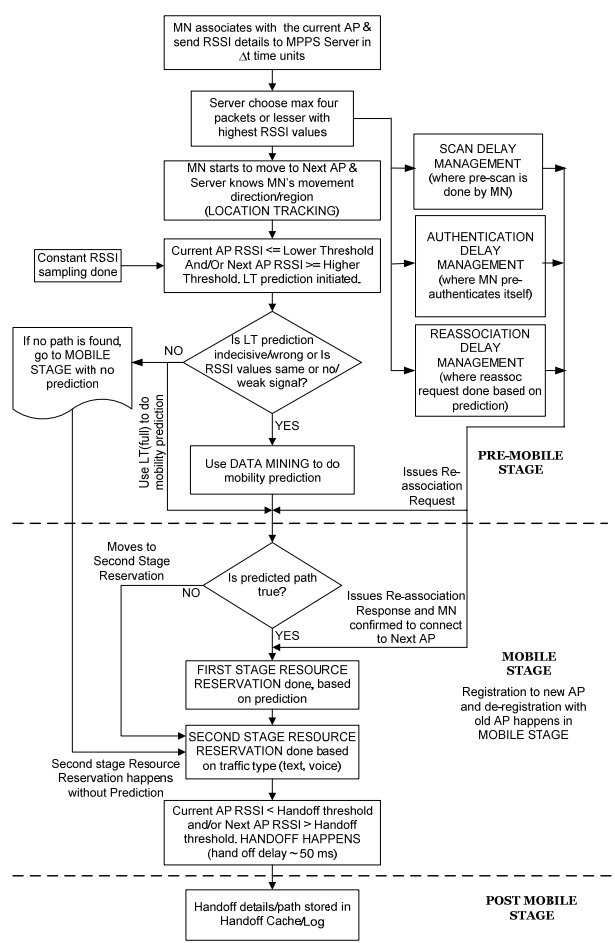

Figure 8. Overall Flow Chart of the Predictive Mobility Management Scheme

In pre-mobile stage, the MN starts moving to next AP and the prediction server knows the current region of MN based on RSSI signals transmitted by MN. In this stage, the three delay managements would be initiated. Scan/Probe delay management happens when the MN pre-scans in regular intervals to know its neighbouring APs RSSI strength and transmission channel. This cuts down scan or probe delay in the next move of MN. Authentication delay management happens, after the first AP association, when the MN is pre-authenticated to all the current AP neighbours. This cuts down authentication delay in the next move of MN. Reassociation delay happens based on prediction result (if available) and the reassociation request is sent to the next predicted AP.

Once the prediction is confirmed, the possible next AP sends the reassociation response to the MN to welcome it to the new AP. If the RSSI values received is fine and if the MN is sensing a very high RSSI signal, then the next AP prediction happens through location tracking, by just analyzing the RSSI signal strength. But if the RSSI signal is not well received or there are equal sets of values or other indecisive scenarios etc, the MN adopts the data mining to do the next AP prediction.

In the mobile stage, the MN is in the process of getting connected to the new AP and is about to disconnect itself from the old AP. If the prediction is true, first stage resource reservation is done on next AP (say, with 5% of available buffer). If the prediction is wrong, first-stage reservation does not happen. The MN in parallel, analyzes the traffic type to see where it is text or audio and make the second stage resource reservation with priority to audio traffic (say, with another 5% of the available buffer). The traffic from the old AP would be buffered into the new AP and the handoff happens, with resource reservation.

In the post-mobile stage, the handoff decision or mobility path is stored in the hand-off cache or log file for future reference. That becomes part of the mobile node path history which can be later used for prediction using data mining.

### A. Proposed Overall Mobility Prediction Algorithm

The overall mobility prediction algorithm is stated below that includes location tracking and data mining.

(1) Initially the Mobile Node (MN) starts by associating with a random Access Point (AP).
(2) The MN transmits the RSSI values received and the surrounding AP's identity (maximum of 4) to MPPS Server. MN stays with that AP for $t$ time units, where $t$ is variable. To be precise, MN sends a packet with details like <RSSI_NextAP, NextAP identity (name, channel no.), RSSI_CurrentAP, CurrentAP identity (name, channel no.), MN identity> value to Server for all the surrounding APs, where RSSI_NextAP and RSSI_CurrentAP are the RSSI values of neighbouring AP and currently attached AP to MN respectively. As each region can have maximum 4 APs and minimum 2 (or 1) APs, the right AP identity is ensured by checking the RSSI value. This transmission happens in regular intervals; say, $\Delta t$ seconds/time units.
(3) For a given MN identity, the server would choose maximum four packets or lesser with the highest RSSI values (for next AP) and note the AP identity (even if the RSSI values match). In the four maximum RSSI values selected, if all the values are above the next AP RSSI threshold value, then the MN is in the inner region with four neighbouring APs, or if only two values are above the next AP RSSI threshold value, then the MN is in the outer most regions with two neighbouring APs, or if only one value is above the next AP RSSI threshold value, then the MN is in the corner region with only one neighbouring AP. Thus, analyzing the next AP RSSI values by the Server would indicate to which AP the MN is getting closer. If one of the RSSI value is highest then the MN is closer to the corresponding AP, or if two of the highest RSSI values (or block values) are the same, then the MN is equally closer to those two APs, or if three or four of the RSSI values are the same, then the MN is in the center of a given region.

(4) After $t$ time units, the MN starts moving and it transmits RSSI sample to MPPS Server. Now the direction of movement of MN can be known to MPPS server, through the RSSI sample values that it receives after $\Delta t$ time units, as explained before. If "$n$" packets sent by the MN are same (having RSSI values same or closer values), then the server knows that the MN is stationary or moving closer to a specific AP for a period of time.

(5) The MPPS server adopts data mining (in parallel) to predict the next AP of MN attachment – when the RSSI sample values received in a set are similar (say, for two or more APs), when the RSSI samples are weak or no signals are coming from neigbouring APs or there is no certain decision from location tracking. Here it uses the mobile node's past history to extract mobility patterns, whereby it can come to a good judgment as explained before. Once the MN moves from one region (current region) to the other region (next region), we need to apply the formula: *Current AP {all neighbor APs} ∩ Next Region {all APs} = {S}*, where S is the set of probable next APs. If S contains only one element, then that is the next predicted AP or if it contains 2 or more elements, it adopts data mining to get the best possible path.

(6) The MPPS server intimates the selected AP (selected with location tracking or selected through data mining) when the RSSI handoff threshold warning is issued that a particular MN is trying to attach to it. It also informs the next AP about initiating the reassociation process and the resources needed to be reserved during the first-stage reservation. A copy of this notice can be sent to MN, optionally. In the next Δt time units, if the MN senses that the current AP RSSI value is lower than the RSSI lower threshold value and the next AP RSSI value is equal to or greater than the RSSI higher threshold value, the MN sends a second-stage reservation to the AP, according to the type of data traffic like text, voice etc. The voice (or video) would imply more buffer space in the second stage reservation request, unlike data that needs fewer buffers.

(7) Handoff can be initiated through RSSI sample comparison. Initially when the RSSI values crosses the RSSI threshold value (lower or higher bound), a RSSI threshold warning is initiated. If the MN senses that the current AP RSSI value is equal to or lesser than the RSSI handoff lower threshold value and/or the next AP RSSI value is equal to or greater than the RSSI handoff higher threshold value, handoff can be initiated from current AP to next AP through MPPS server. During the handoff, the current AP's permission is sought to allow MN to get connected to next AP, which is the reassociation process.

(8) Once the handoff is successful, the mobile node path is stored in the MPPS cache or logged into a central database.

Note: If the MN at any point senses that it is losing the network connection from the current AP, because of some delay in the communication from the MPPS server or so, the MN can initiate the handoff to the new AP. This can be a backup option.

## VI. PERFORMANCE SIMULATION RESULTS FOR MOBILITY PREDICTION

The mobility model used in Java simulation is a modified random waypoint model, where mobile nodes move randomly to the sides and to the center. The mobile node travels in a 6 x 6 region matrix which contains 5 x 5 access point matrix. Thus the 36 regions have 25 access points placed in a matrix format as shown in fig. 1.

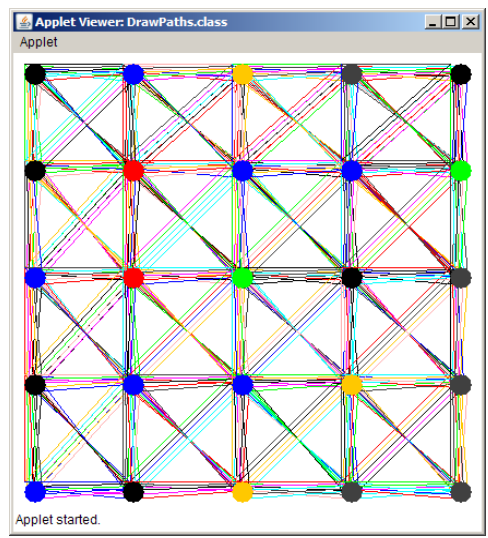

Figure 9. Mobile Node Connections generated (showing around 10,000 paths, some of which are overlapped in the applet)

The simulation is run to create 10,000 mobile paths and this is stored in a central database of MPPS server as shown in fig. 9, which displays the uniformity of mobile node's path connections. This data set is used for data mining and to create the mobility patterns. Later, a test set of 10 random mobile paths were created (with a maximum of six access points and a minimum of three access points) and tested for prediction accuracy. The accuracy of prediction through RSSI sample approach and location tracking is 100%, if the RSSI value received by MN is high from a particular AP, during its mobile path. The data mining approach which was done in parallel was simulated and the prediction analysis is given below.

We would like to show the prediction accuracy for a sample of 30 random mobile node paths, as given below.

The mobile node movement history or cache consisted of 10,000 mobile paths as follows.

Path 1:  6(7) → 0(0) → 1(1)
Path 2:  14(16) → 8(9) → 2(2) → 1(1) → 0(0) → 5(6)
Path 3:  16(19) → 21(25) → 15(18)
Path 4:  10(12) → 5(6)
Path 5:  3(3) → 2(2) → 1(1) → 0(0) → 6(7)
Path 6:  21(25) → 15(18) → 10(12)
Path 7:  12(14) → 6(7) → 0(0) → 1(1)
Path 8:  7(8) → 1(1) → 0(0) → 6(7)
Path 9:  4(4) → 3(3) → 2(2) → 1(1) → 6(7) → 0(0)
Path 10: 7(8) → 8(9) → 2(2) → 1(1)
Path 11: 1(1) → 0(0) → 5(6)
Path 12: 18(21) → 17(20) → 16(19) → 15(18) → 10(12)
Path 13: 4(4) → 3(3) → 2(2) → 1(1) → 0(0) → 5(6)
Path 14: 17(20) → 16(19) → 15(18) → 10(12)
Path 15: 9(10) → 8(9) → 3(3) → 2(2) → 1(1) → 0(0)
Path 16: 7(8) → 6(7) → 5(6) → 0(0)
Path 17: 16(19) → 15(18) → 10(12)
Path 18: 13(15) → 12(14) → 11(13) → 10(12) → 5(6)
Path 19: 5(6) → 0(0)
Path 20: 14(16) → 13(15) → 12(14) → 11(13) → 10(12) → 5(6)
Path 21: 7(8) → 6(7) → 1(1) → 0(0)
Path 22: 13(15) → 14(16) → 19(22) → 18(21) → 17(20)
Path 23: 15(18) → 10(12)
Path 24: 6(7) → 5(6) → 0(0)
Path 25: 9(10) → 4(4) → 3(3) → 2(2) → 1(1) → 0(0)
Path 26: 22(26) → 16(19) → 15(18) → 20(24)
Path 27: 1(1) → 0(0) → 5(6)
Path 28: 0(0) → 1(1)
Path 29: 13(15) → 7(8) → 6(7) → 1(1) → 0(0)
Path 30: 22(26) → 21(25) → 20(24) → 15(18)

The values in the parentheses are the region numbers. i.e. 7(8) means AP7 in region 8. As a result of the simulation, the table I, II and III shows the prediction accuracy for the random sample considered above, using three different prediction schemes. The average accuracy of LTDMPS (with data mining and partial location tracking) was 91% for 30 paths as in table I.

But the overall LTDMPS prediction accuracy for 10,000 mobile paths was found to be 81.95%. The predicted path contains, for example entries like 2[2, 1] in row 2 of table I. The 2 is the predicted next AP and inside the square brackets, 2 is the actual node to which MN moved and 1 is the frequency rank of the predicted node, which is AP2. We compared our prediction method with two other methods, as baseline schemes. Refer to tables II and III.

The first prediction method is called the Transition Matrix (TM) prediction scheme. In this method, a cell-to-cell transition matrix is formed by considering the previous inter-cell movements of the mobile nodes or users. The predictions are based on this transition matrix by selecting the *x* most likely cells or regions as the predicted cells. We used TM for performance comparison because it makes predictions based on the previous movements of the mobile node or user [22]. Assuming x=1 (as the previous scheme also used x=1), the average accuracy of TM was found to be 60% for 30 paths as in table II. But the overall average TM prediction accuracy for 10,000 mobile paths was found to be 52.49%.

TABLE I. PREDICTED PATH ACCURACY TABLE FOR A SAMPLE USING LTDMPS

| Predicted Path  (Predicted AP[Original AP, Rank]) | Accuracy (%) |
|---|---|
| 6[6]→0[0,1]→1[1,1] | 100% |
| 14[14]→8[8,1]→2[2,1]→1[1,1]→0[0,1]→5[5,1] | 100% |
| 16[16]→15[21,2]→20[15,2] | 0% |
| 10[10]→5[5,1] | 100% |
| 3[3]→2[2,1]→1[1,1]→0[0,1]→5[6,2] | 75% |
| 21[21]→15[15,1]→10[10,1] | 100% |
| 12[12]→6[6,1]→0[0,1]→1[1,1] | 100% |
| 7[7]→1[1,1]→0[0,1]→5[6,2] | 68.7% |
| 4[4]→3[3,1]→2[2,1]→1[1,1]→0[6,2]→5[0, 2] | 60% |
| 7[7]→2[8,2]→3[2,2]→1[1,1] | 33.3% |
| 1[1]→0[0,1]→5[5,1] | 100% |
| 18[18]→17[17,1]→16[16,1]→15[15,1]→10[10,1] | 100% |
| 4[4]→3[3,1]→2[2,1]→1[1,1]→0[0,1]→5[5,1] | 100% |
| 17[17]→16[16,1]→15[15,1]→10[10,1] | 100% |
| 9[9]→8[8,1]→3[3,1]→2[2,1]→1[1,1]→0[0,1] | 100% |
| 7[7]→6[6, 1]→5[5, 1]→0[0,1] | 100% |
| 16[16]→15[15,1]→10[10,1] | 100% |
| 13[13]→12[12,1]→11[11,1]→10[10,1]→5[5,1] | 100% |
| 5[5]→0[0, 1] | 100% |
| 14[14]→13[13,1]→12[12,1]→11[11,1]→10[10,1]→5[5,1] | 100% |
| 7[7]→6[6,1]→1[1,1]→0[0,1] | 100% |
| 13[13]→8[14,2]→13[19,2]→18[18,1]→17[17,1] | 50% |
| 15[15]→10[10,1] | 100% |
| 6[6]→5[5, 1]→0[0, 1] | 100% |
| 9[9]→4[4,1]→3[3,1]→2[2,1]→1[1,1]→0[0,1] | 100% |
| 22[22]→16[16,1]→15[15,1]→20[20,1] | 100% |
| 1[1]→0[0,1]→5[5,1] | 100% |
| 0[0]→1[1,1] | 100% |
| 13[13]→7[7,1]→6[6,1]→1[1,1]→0[0,1] | 100% |
| 22[22]→21[21,1]→20[20,1]→15[15,1] | 100% |

The second prediction method is the Ignorant Prediction (IP) scheme [23]. This approach disregards the information available from movement history. To predict the next inter-cell movement of a user, this method assigns equal transition probabilities to the neighboring cells of the mobile node's currently resident cell. It means that prediction is performed by randomly selecting *m* neighboring cells of the current cell. We have taken *m* to be the maximum no. of neighbours possible. The value in the parenthesis in the paths shows the corrected AP number. The average accuracy of this scheme was found to be 21% and as expected, was quite inconsistent. But the overall average IP prediction accuracy for 10,000 mobile paths was found to be 19.3%.

A comparison line graph in fig. 10 shows the prediction accuracy for the three different schemes, for the 30 mobile paths discussed before.  It is very clear that LTDMPS (our proposal) which averages 89.5% prediction accuracy for 30 paths is much better compared to TM (60.5% accuracy) and IP (17.4% accuracy) schemes.

TABLE II. PREDICTED PATH ACCURACY TABLE FOR A SAMPLE USING TRANSITION MATRIX (TM)

| Predicted Path (Predicted AP[Original AP, Rank]) | Accuracy (%) |
|---|---|
| 6(6)→5[0, 3]→5[1, 3] | 0% |
| 14[14]→9[8, 3]→3[2, 3]→1[1, 1]→0[0, 1]→5[5, 1] | 60% |
| 16[16]→15[21, 4]→20[15, 3] | 0% |
| 10[10]→5[5, 1] | 100% |
| 3[3]→2[2, 1]→1[1, 1]→0[0, 1]→5[6, 2] | 75% |
| 21[21]→20[15, 3]→10[10, 1] | 50% |
| 12[12]→7[6, 3]→5[0, 3]→5[1, 3] | 0% |
| 7[7]→2[1, 3]→0[0, 1]→5[6, 2] | 33.3% |
| 4[4]→3[3, 1]→2[2, 1]→1[1, 1]→0[6, 2]→5[0, 3] | 60% |
| 7[7]→2[8, 5]→3[2, 3]→1[1, 1] | 33.3% |
| 1[1]→0[0, 1]→5[5, 1] | 100% |
| 18[18]→13[17, 2]→16[16, 1]→15[15, 1]→10[10, 1] | 75% |
| 4[4]→3[3, 1]→2[2, 1]→1[1, 1]→0[0, 1]→5[5, 1] | 100% |
| 17[17]→16[16, 1]→15[15, 1]→10[10, 1] | 100% |
| 9[9]→4[8, 2]→3[3, 1]→2[2, 1]→1[1, 1]→0[0, 1] | 80% |
| 7[7]→2[6, 2]→5[5, 1]→0[0, 1] | 66.7% |
| 16[16]→15[15, 1]→10[10, 1] | 100% |
| 13[13]→8[12, 2]→7[11, 2]→10[10, 1]→5[5, 1] | 50% |
| 5[5]→0[0, 1] | 100% |
| 14[14]→9[13, 2]→8[12, 2]→7[11, 2]→10[10, 1]→5[5, 1] | 40% |
| 7[7]→2[6, 2]→5[1, 2]→0[0, 1] | 33.3% |
| 13[13]→8[14, 5]→9[19, 3]→14[18, 2]→13[17, 2] | 0% |
| 15[15]→10[10, 1] | 100% |
| 6[6]→5[5, 1]→0[0, 1] | 100% |
| 9[9]→4[4, 1]→3[3, 1]→2[2, 1]→1[1, 1]→0[0, 1] | 100% |
| 22[22]→21[16, 3]→15[15, 1]→10[20, 2] | 33.3% |
| 1[1]→0[0, 1]→5[5, 1] | 100% |
| 0[0]→5[1, 3] | 0% |
| 13[13]→8[7, 3]→2[6, 2]→5[1, 2]→0[0, 1] | 25% |
| 22[22]→21[21, 1]→20[20, 1]→15[15, 1] | 100% |

TABLE III. PREDICTED PATH ACCURACY TABLE FOR A SAMPLE USING IGNORANT PREDICTION

| Predicted Path (Predicted AP[Original AP]) | Accuracy (%) |
|---|---|
| 6[6]→2[0]→5[1] | 0% |
| 14[14]→8[8]→12[2]→6[1]→6[0]→6[5] | 20% |
| 16[16]→15[21]→15[15] | 50% |
| 10[10]→11[5] | 0% |
| 3[3]→4[2]→3[1]→7[0]→5[6] | 0% |
| 21[21]→17[15]→20[10] | 0% |
| 12[12]→6[6]→12[0]→1[1] | 66.7% |
| 7[7]→13[1]→7[0]→5[6] | 0% |
| 4[4]→3[3]→9[2]→7[1]→6[6]→2[0] | 40% |
| 7[7]→3[8]→3[2]→6[1] | 0% |
| 1[1]→5[0]→6[5] | 0% |
| 18[18]→23[17]→22[16]→21[15]→11[10] | 0% |
| 4[4]→8[3]→4[2]→6[1]→6[0]→5[5] | 20% |
| 17[17]→16[16]→10[15]→16[10] | 33.3% |
| 9[9]→4[8]→2[3]→7[2]→1[1]→0[0] | 40% |
| 7[7]→8[6]→12[5]→6[0] | 0% |
| 16[16]→15[15]→21[10] | 50% |
| 13[13]→17[12]→8[11]→5[10]→6[5] | 0% |
| 5[5]→11[0] | 0% |
| 14[14]→13[13]→9[12]→7[11]→16[10]→16[5] | 20% |
| 7[7]→6[6]→7[1]→5[0] | 33.3% |
| 13[13]→9[14]→19[19]→14[18]→17[17] | 50% |
| 15[15]→10[10] | 100% |
| 6[6]→11[5]→6[0] | 0% |
| 9[9]→3[4]→8[3]→9[2]→6[1]→5[0] | 0% |
| 22[22]→18[16]→17[15]→20[20] | 33.3% |
| 1[1]→5[0]→1[5] | 0% |
| 0[0]→6[1] | 0% |
| 13[13]→18[7]→8[6]→7[1]→7[0] | 0% |
| 22[22]→17[21]→20[20]→15[15] | 66.7% |

The division of surrounding AP area into four regions plays a great role in making the proposal more accurate. This would reduce the number of possible next AP set as the MN moves to a new region and hence the prediction can be more accurate. If such a division is not made, then based on the network model used in fig. 1, there can be eight neighbours to an AP, and to predict one of them would be quite a difficult task. So the location tracking (partial) through RSSI samples, coupled with data mining (full) makes our proposal work better. Note the thick dots that touch 100% a few times with LTDMPS.

Another observation that can be made is the frequency rank of the nodes (shown in brackets) in the predicted path for LTDMPS and TM from table I and II. From table I for LTDMPS, it is clear that the frequency rank is mostly 1, 2 or 3 (seen in other samples). Frequency rank 1 means prediction is correct. But from table II for TM, the frequency ranks fluctuates between 1, 2, 3, 4 and 5, thus showing how it's lacking in accuracy. The maximum frequency deviation graphs for LTDMPS and TM schemes are shown in fig. 11 and fig. 12 respectively.

As per fig. 11 and fig. 12, the bars that correspond to frequency rank 1 are the right prediction. Other ranks show that during wrong prediction, the rightly predicted access point is ranked with that number. For LTDMPS (our scheme), even in the wrongly predicted case, the rank of the right access point is 2.

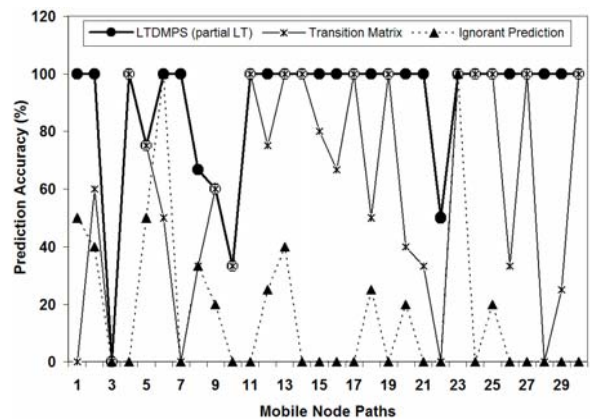

Figure 10. Prediction accuracy graph for 30 random sample paths for LTDMPS (partial LT and DM), TM and IP Schemes

It can be maximum 3 also as per other samples in 10,000 paths. This is because of the concept of region that we have in our proposal.

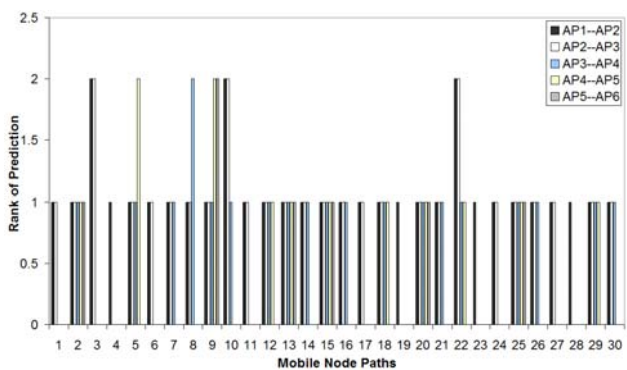

Figure 11. Maximum Frequency Deviation graph for 30 random sample paths for LTDMPS

Consider the 30 mobile node paths used for analysis considered before. We ran the simulation by using only location tracking and noted that the mobility prediction accuracy was 44.8%. We purposely injected errors to make the accuracy below 50%. As seen before, the mobility prediction using data mining with partial location tracking gave 89.5% prediction accuracy.

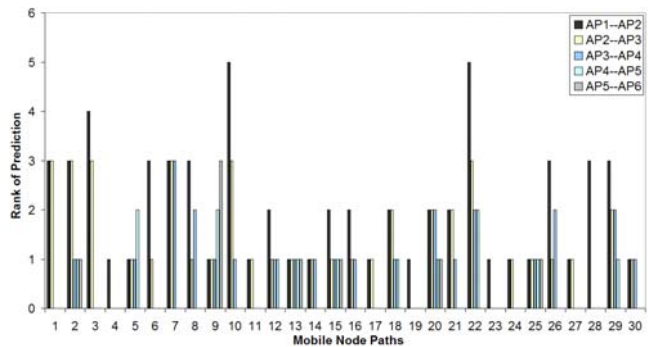

Figure 12. Maximum Frequency Deviation graph for 30 random sample paths for TM

So the combined prediction accuracy (using both schemes, as independent events) under simulation was found to be around 98.5%, which is quite a good result.

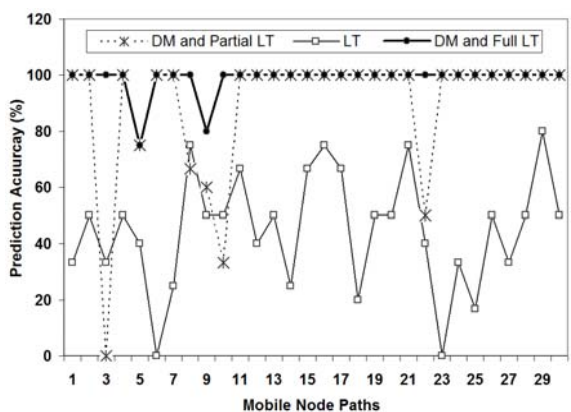

Figure 13. Prediction accuracy graph for 30 random sample paths for LTDMPS (using partial LT and DM), Location Tracking, LTDPMS (using full LT and DM)

This is because it would take the best (i.e. right) prediction in both cases. The graph of such a scenario would be as in fig. 13 which shows the above mentioned comparisons. The overall average accuracy of the combined scheme of LTDMPS (with full location tracking and data mining) for 10,000 mobile paths was calculated to be 92.7%. The flow chart of the combined mobility prediction process is shown in fig. 14. The diagram shows the timing of the mobility prediction.

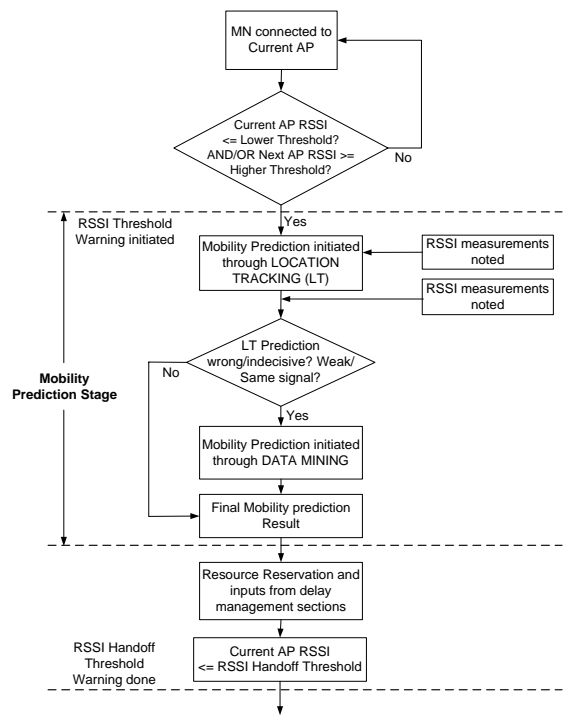

Figure 14. The combined mobility prediction process

The location tracking prediction is initiated just after the RSSI threshold warning. The prediction moves to data mining mode in the following two cases: (1) if the RSSI signals are weak and location tracking is not possible or equal, where location tracking cannot choose a final next AP; (2) if the location tracking prediction is wrong, as MN is constantly sampling the RSSI values through the MPPS server. In this mobility management scheme, handoff of an MN over to a new AP happens only if the current signal level drops below the RSSI handoff lower threshold and/or the next AP is stronger (RSSI handoff higher threshold) than the current one by a given hysteresis margin. Preference is always given to the dropping of RSSI value of current AP below the RSSI handoff lower threshold.

## VII. RELATED WORKS ON DELAY AND RESOURCE MANAGEMENT

Handoff delay measured for different client wireless adapter and access points from different vendors as per [13], [24] are as shown in table IV. From the table, we can find the average delay values to be as follows: Average Probe delay = 238.52 msec; Average Authentication delay = 2.12 msec; Average Reassociation delay = 3.25 msec. It clearly shows that the probe delay is the biggest of all delays. It is well known that for running multimedia applications in a wireless network effectively, the average handoff delay should be less than or around 50 msec [25]-[26].

TABLE IV. HANDOFF DELAYS IN DIFFERENT WIRELESS PRODUCTS

| AP MN | Cisco | | | Soekris | | | Lucent | | |
|---|---|---|---|---|---|---|---|---|---|
| | P | A | R | P | A | R | P | A | R |
| Lucent | 37.2 | 3.08 | 5.07 | 196.9 | 1.77 | 1.77 | 81.2 | 0 | 1.73 |
| Cisco | 399.8 | 3.56 | 4.13 | 349.9 | 4.48 | 1.79 | 347.3 | 1.49 | 1.09 |
| ZoomAir | 195.6 | 2.40 | 8.84 | 191.3 | 2.37 | 1.77 | 347.5 | 0 | 3.09 |

Legend (all in msec): P – Probe Delay; A – Authentication Delay; R – Reassociation Delay

There are two different approaches available for reducing the probe delay in 802.11 networks. They are, limiting the number of channels to be probed, and reducing the waiting time for each channel to be probed.

The probe delay is dependant on the scan mode used (i.e. passive scan or active scan). Probe delay under passive scan would be a function of the beacon interval and the number of channels available. So for IEEE 802.11b with 11 channels and 100 msec beacon interval, the probe delay would be 1100 msec. The channel switching delay is very minimal and of the order of 40 to 150 μsec [27]. The probe delay through active scan can be controlled by using optimized values for *MaxChannelTime* and *MinChannelTime* variables. So the active scan probe delay ($T_{pd}$) can be expressed as $N \times MinChannelTime <= T_{pd} <= N \times MaxChannelTime$, where N is the available number of channels. If no probe response is received by *MinChannelTime* in a channel, the mobile node probes the next channel. If one or more probe responses are received by *MinChannelTime*, the MN stops accepting probe responses at *MaxChannelTime* and processes all received responses. Velayos et al. proposed the tuning scheme that finds more appropriate or optimal *MaxChannelTime* and *MinChannelTime* values to reduce the channel probing delay [28].

Shin et al. presented an innovative discovery method [15] using the neighbor graph (NG) and non-overlap graph (NOG). This scheme (referred to as the NG-pruning scheme) focuses on reducing both the total number of channels to be probed and the waiting time on each channel. They suggested two algorithms: the NG and NG pruning algorithms. The rationale behind these algorithms is to ascertain whether a channel needs to be probed or not (by the NG algorithm) and whether the MH has to wait more probe response messages on a specific channel before the expiration of *MaxChannelTime* (by the NG-pruning algorithm). The non-overlap graph in the NG-pruning scheme allows the channel waiting time to be shortened as the mobile node need not wait for further probe responses once it receives some probe responses from existing APs.

In the channel mask scheme [29], only a well-defined subset of all available channels is probed. Channel selection is performed by means of a channel mask that is preloaded when the mobile node starts up. A full-scan is done initially and the channel mask is then constructed by the information gathered in the first full-scan. In IEEE 802.11b, only three channels do not overlap among all 11 channels. Hence, in a well configured wireless network, all or most of the APs operate on channels 1, 6, and 11. Consequently, the channel mask is formed by combining three frequent channels (i.e., 1, 6, and 11) and the channels scanned at the first full-scan [24].

SyncScan [16] and MultiScan [17] also reduce the number of channels to be probed. But, they are not dependent on the empirical handoff information and collect information on nearby APs through continuous monitoring. When IEEE 802.11 wireless networks are installed in public places, secure authentication becomes quite a critical issue. However, more secure authentication schemes will increase the overhead and results in longer authentication and/or re-association delays. So reducing the authentication and reassociation delay is as important as reducing the probe delay in such a case. In another work, Pack has proposed a predictive authentication scheme by using a modified version of IEEE 802.1x key distribution model. Here a Frequent Handoff Region (FHR) is defined which is the adjacent APs around a mobile node [14].

A proactive scheme called the proactive neighbor caching (PNC) scheme based on a distributed cache structure was introduced in [30]. It uses a neighbor graph, which dynamically captures the mobility topology of a wireless network. The PNC scheme ensures that the mobile node's context is always dispatched one-hop ahead, and therefore the handoff delay can be greatly reduced. Here, the context includes information about the mobile node's session, the quality of service (QoS), and security details [31].

In the FHR scheme, the central system constructs the frequent handoff region by considering the mobility history and profile of the mobile nodes. On the contrary, in the PNC scheme, each AP learns the mobility patterns of the mobile nodes and configures the neighbor graph in a distributed manner [24].

Fathi et al. discussed the impact of security on latency in WLAN 802.11b and finds that the major contributor of the authentication delay is the probing time needed to detect the surrounding access point [32]. In their paper, they proposed to evaluate the overhead introduced by the security mechanisms in WLAN such as authentication, by developing an analytical model based on random errors to evaluate the authentication delay for various error rates and also measure the authentication delay for WLAN 802.11b using CISCO Access Point and client cards. Braun et al. reported an efficient authentication and authorization of mobile users based on peer-to-peer network mechanisms [33].

Aura et al. had proposed reducing re-authentication delay in wireless networks where in their paper to reduce the authentication delay and to enable optimistic access without opening a window for fraudulent access [34]. They presented a protocol for the re-authentication of a mobile node when it repeatedly connects to different access points.

Ferng et al. has proposed a dynamic resource reservation scheme with mobility prediction for wireless multimedia networks [35]. To reduce unnecessary resource reservation, the prediction of moving direction can be incorporated to enhance the pure resource reservation. Therefore, a dynamic resource reservation scheme with mobility prediction is proposed where a resource request can be done with available range rather than a fixed resource request is specified by each traffic flow to make the reservation scheme flexible. Zhu et al. has proposed resource reservation for handoff using location-aided mobility prediction for wireless cellular networks [36]. They analyze the performance of the proposed scheme and proposed a two-dimensional random walk model for performance evaluation.

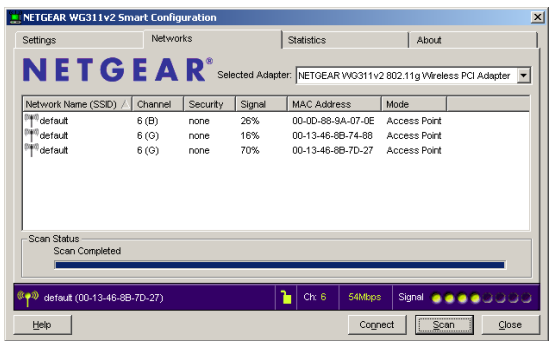

Figure 15. The mobile node's association with access point that provides highest signal

Generally, the mobile node would attach itself to the access point which provides the highest signal as shown in fig. 15. Here out of three access points on channel 6, the mobile node connects to the access point (00-13-46-8B-7D-27) that provides 70% signal strength, compared to the ones with 26% and 16% signal strengths.

## VIII. DELAY AND RESOURCE RESERVATION MANAGEMENT

Once the mobility prediction is done, there are two specific processes to make the MN handoff smooth as much as possible. One is delay management and the other is the resource reservation management.

### A. Delay Management

There are three delays that a mobile node faces as it moves around from one AP to another – Probe, Authentication and Reassociation delays as shown in fig. 16. These delays play an important role during the process of MN handoff to a new AP. Reducing these delays would hasten and smoothen the handoff process. The process is explained below.

(1) The access points would be sending information about itself (identity, transmission channel etc) to the MPPS server and updates the server if there is any change. Once a MN attaches itself to the first AP, the server updates the MN about its neighbours, through the AP of attachment. MN that is located in a region does active multi-channel scanning (based on information from server) where it sends probe requests for each channel scanned to locate the neighbour APs and notes their AP's RSSI values along with the channel of transmission. It sends this information to the MPPS server. Once the MN moves, it already knows the channel where it located the next AP and avoids unnecessary channel scans. The central MPPS server can also update the MN (in case of any change or lack of information), as the server is constantly updated of an AP's channel of transmission. This would reduce the probe delay significantly.

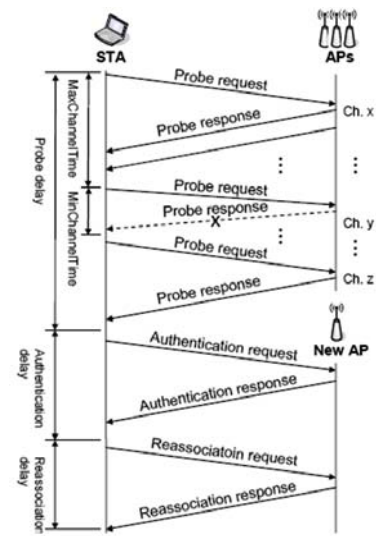

Figure 16. IEEE 802.11 handoff process and the delays involved when a mobile node moves across access points (Ping-Jung *et al*., 2006)

(2) When the MN gets connected to any AP, it would initiate authentication with all the neighbouring APs, horizontally, vertically and diagonally. Later when it moves to next AP, there would be no authentication delay as the MN is already authenticated. Once connected to the new AP, the MN again initiates authentication with all the neighbouring APs, if it's not pre-authenticated with that. This would reduce the authentication delay greatly.

(3) Once the next AP prediction is known (based on the prediction algorithm discussed before), the process of reassociation is initiated with the *confirm flag signal* as false. This will initiate the process but will not confirm it. Once the MN or MPPS server knows that the prediction is true, the flag signal is sent to the new AP as true (as a piggybacked communication), and only then the reassociation is confirmed. The confirmed flag signal if not communicated as true (within a time duration) will ensure that the prediction is wrong and that would cancel the reassociation process initiated. Some timer can be set for the cancellation purposes. This would reduce the reassociation delay.

*B. Resource Reservation Management*

Using the mobility prediction scheme, once the next AP is predicted, steps for resource reservation management can be done. There can be two stages in the reservation process – first stage and second stage reservations.

*First Stage Reservation:* In the first stage reservation, once the next AP prediction is done through location tracking and data mining, the buffers are blocked (say, around 5% free buffer space) on it explicitly through the MPPS server. Later as the MN moves toward a particular AP and when RSSI value is greater than the RSSI higher threshold value, the reservation confirmation can be given through *a confirm flag signal*, during the mobile node association (say, as a piggybacked communication).

*Second Stage Reservation:* Based on the Type of Service or ToS field (in IPv4 packets) or Traffic Class field in combination with Flow Label field (in IPv6 packets) the packets in a network can be classified. The ToS field helps the host to specify a preference for how the datagram would be handled as it goes through the network. Some specific application protocols for audio and video can be given higher priority than the other text based protocols/applications. In the second stage reservation, when the MN sees the packets that is going through it to be of a particular type, that information is communicated by MN to the MPPS server which in turn notifies the next access point and some buffer space (say, around 5% free buffer space) is marked for reservation. In a sense, only blocking of buffer space is done, which can be released for other emergencies, if any. Some fine tuning can be done here, by choosing the next AP with *n* higher ranks. The actual reservation will only be done when a confirmation is given later, through *a confirm flag signal,* during the mobile node association (say, as a piggybacked communication).

## IX. PERFORMANCE SIMULATION RESULTS FOR DELAY AND RESOURCE MANAGEMENT

Simulation done here also uses a mobility model that is a modified version of Random waypoint model, where the mobile nodes move uniformly to the sides and to the center. As before, the mobile node travels in a 6 x 6 region matrix which contains 5 x 5 access point matrix. Thus the 36 regions have 25 access points placed in a matrix format. The performance graphs shown are based on the estimates from existing real world estimates that were compared with the proposed mobility management solution through Java simulation and shows the anticipated results. The simulation parameters are shown in table V.

We used Java package edu.cornell.lassp.houle.RngPack with uniform pseudo-random number generators to generate random numbers. RANMAR, a lagged Fibonacci generator proposed by Marsaglia and Zaman is used [37] We initialized the wireless interface with parameters (as shown in table V) to make it work like the 914MHz Lucent WaveLAN DSSS radio interface (taken from ns-2 file).

Also, Friis free space equation used for RSSI calculation [38] is used as follows in equation 1:

$$P_r(d) = \frac{P_t * G(r) * \lambda^2}{(4*\pi)^2 * d^2 * L} \quad (1)$$

where, $P_r(d)$ is the received power, $P_t$ is the transmitted power, $G(t)$ and $G(r)$ are transmitter and receiver antenna gains, $\lambda$ is the wavelength, d is the T-R separation in meters, and L is the system loss factor not related to propagation (>=1).

Queuing theory is used to calculate mean delay experienced by packets [38] as follows in equation 2:

$$T = \frac{1}{a} + \frac{1}{\left[1 - \frac{b}{a}\right]} \quad (2)$$

where, *a* and *b* are the mean processing capacity and the mean packet arrival rate, respectively.

TABLE V. SIMULATION PARAMETERS

| Simulation Variables | Values |
|---|---|
| buffer_size | 100 MB |
| audio_file_size | 20 MB |
| text_file_size | 5 MB |
| minChannelTime | 2 msec |
| maxChannelTime | 6 msec |
| pingTime | 3 msec (low load) |
| onewayTime | 2 msec (low load) |
| RSSImax | 100 mW |
| RSSIthreshold | 3 mW to 4 mW |
| RSSIhandoff | 1 mW to 2 mW |
| antenna_height | 1.5 m |
| wireless_frequency | 914e+6 Hz |
| bandwidth | 2*1e6 Hz |
| receivepower_threshold | 1.427e-08 W (for 100m) |
| trans_power (transmitted power) | 100 mW |
| proc_cap (processing capacity) | 1000000 packets |
| arrivalrate_max | 950000 packets/sec |
| arrivalrate_avg | 650000 packets/sec |
| arrivalrate_min | 150000 packets/sec |
| loadMax | >=10 nodes |
| loadAvg | >=5 nodes && <10 nodes |
| loadMin | >=1 node && <5 nodes |

In the existing technology, for passive scan, if the beacon interval is 100 msec, the average probe delay of IEEE 802.11b with 11 channels is 1100 msec and 802.11a with 32 channels is 3200 msec. This indicates that to probe one channel, it is taking around 100 msec. However, we are not using passive scan in our scheme [24].

For active scan, in 802.11b and 802.11a networks, all the channels would be probed to look for the next AP. This can be lesser when only few non-overlapping channels are probed (like channel 1, 6 and 11) or if configured that way. In our new proposal, the MN gets to know all its neighbour AP (in a given region) and the channel they are transmitting, which enables it to cut down the probe delay completely as the AP identity and channel number is already known. A normal ping or probe request from mobile node would confirm the AP's presence and that would take on the average 3 msec to 5 msec, depending on load. Except for the initial scan when a mobile node enters the wireless network, the probe delay later can be minimized to less than 10 msec. The probe delay ($T_{pd}$) of mobile node's active scan can be expressed as, $N \times MinChannelTime \leq T_{pd} \leq N \times MaxChannelTime$, where N is the number of channels scanned. The optimal value of *MinChannelTime* (if no response by this time, MN scan next channel) can be around 2 to 4 msec and that of *MaxChannelTime* (MN stops probing by this time, if responses received) can be 6 to 8 msec for the multi-channel active scanning. This can speed things up. Fig. 17 shows the probe delay graph for 10 mobile node paths as listed in table VI. The value averages to 9.8 msec. The average probe delay for 100 paths under simulation was found to be 9.9 msec.

TABLE VI. PREDICTED PATH ACCURACY TABLE (10 PATHS) USING LTDMPS (USING FULL LT AND DM)

| Predicted Path  (Predicted AP[Original AP, Rank]) | Accuracy (%) |
|---|---|
| 6[6]→0[0,1]→1[1,1] | 100% |
| 14[14]→8[8,1]→2[2,1]→1[1,1]→0[0,1]→5[5,1] | 100% |
| 16[16]→15[15,1]→20[20,1] | 100% |
| 10[10]→5[5,1] | 100% |
| 3[3]→2[2,1]→1[1,1]→0[0,1]→5[5,1] | 100% |
| 21[21]→15[15,1]→10[10,1] | 100% |
| 12[12]→66,1]→0[0,1]→1[1,1] | 100% |
| 7[7]→1[1,1]→0[0,1]→5[5,1] | 100% |
| 4[4]→3[3,1]→2[2,1]→1[1,1]→0[0,1]→5[5, 1] | 100% |
| 7[7]→2[2,1]→3[2,2]→1[1,1] | 66.6% |

As mentioned before, in the subsequent scans the scan delay is lessened, as the mobile node has already the knowledge of access points, along with its transmission channel. Based on the network load, other parameters like collision, load delay, transmission rate and other traffic variables would also be affected. In fact, scan delay = $f$ (load, *MinChannelTime*, *MaxChannelTime*).

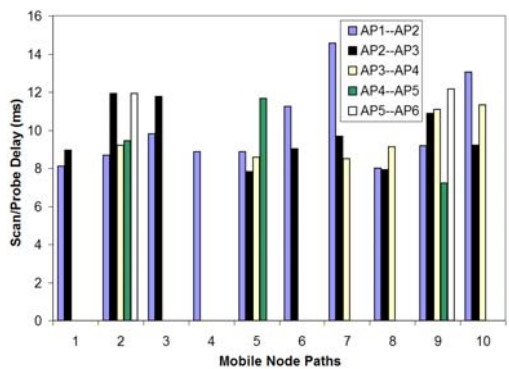

Figure 17. Probe or scan delay graph during handoff showing 10 mobile node paths

Since the mobile node authenticates itself with all the neighbouring access points, the authentication delay is only significant when the MN first enters the wireless network. After that there would be negligible authentication delay (around 5 msec or so), as it is already authenticated to the possible next AP that it is moving to.

The authentication delay includes delay for four message exchanges, namely Challenge-Request, Challenge-Response (AP sends nonce, RN), Response (MN signs RN with WEP encryption), and Approval (AP verifies RN). In fact, authentication delay = $f$ (load, authentication type). In the simulation, the authentication delay was implemented as a function of the network load and shared authentication using WEP for the first time and later as a function of network load only.

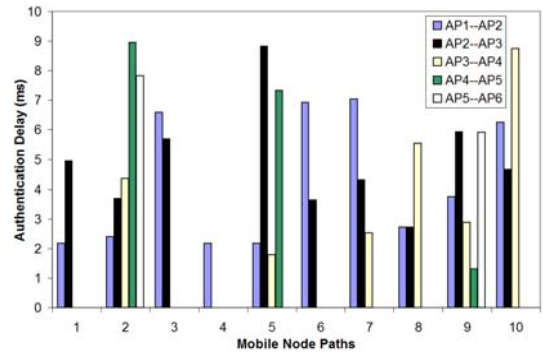

Figure 18. Authentication delay graph during handoff showing 10 mobile node paths

Fig. 18 shows the authentication delay comparison for ten mobile node paths listed in table VI. The value averages to 4.6 msec. As noted before, once the MN gets connected to an AP, all the neighbouring APs are pre-authenticated. So for the next move, there is no need for authentication, only a ping time to check for the access point's existence.  The average authentication delay for 100 paths was found to be 4.3 msec.

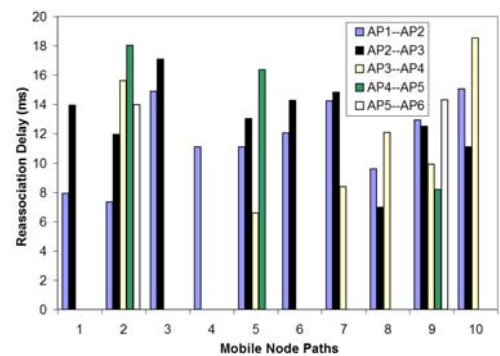

Figure 19. Reassociation delay graph during handoff showing 10 mobile node paths

The reassociation delay would follow the authentication delay that would be around 4 msec for two reassociation frame (reassociation request and response) exchanges and around 10 to 12 msec when IAPP (Inter Access Point Protocol) is implemented, where there would be around six frames (reassociation request and response, IAPP frames like – send security block, ACK security block, Move request, Move response) exchanged [24]. Fig. 19 shows the reassociation delay comparison for ten mobile node paths listed in table VI. The value averages to 12.5 msec. In fact, reassociation delay = $f$ (load, IAPP protocol). In the simulation, the reassociation delay was implemented as a function of the network load and IAPP. The average reassociation delay for 100 paths was found to be 12.2 msec.

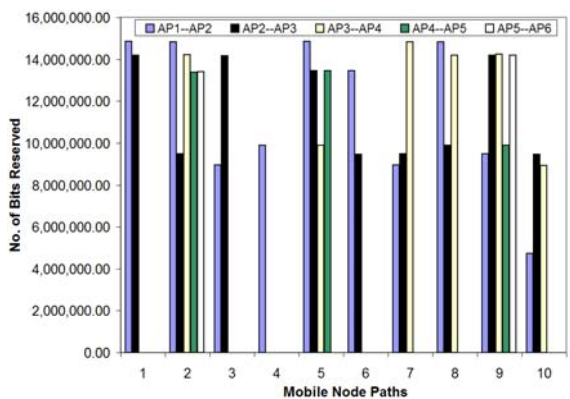

Figure 20. Resource allocation in the proposed scheme for 10 mobile node paths

Fig. 20 shows a constant resource allocation for the proposed mobility management scheme. The default buffer size on access point is taken as 100MB. Voice and text traffic are identified from packets and treated differently. The resource reservation is done in two stages. The first stage reservation (say, 5% of free buffer) is done based on the mobility prediction and second stage reservation (say, 5% of free buffer) is done based on the traffic type. In the simulation, the resource reservation was implemented as a function of prediction result and traffic type (i.e. $f$ (prediction result, traffic type)).

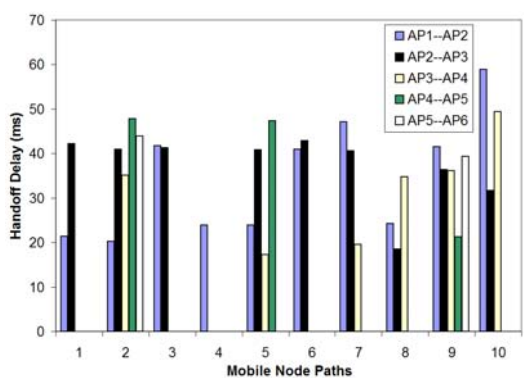

Figure 21. Handoff delay graph with the proposed mobility management architecture for 10 mobile node paths

As shown in fig. 21, the handoff is smoother with the new proposal, as the overall handoff delay is around 50 msec or lesser, which eventually is good for voice communications. The delays are obviously high for the first AP in a mobile path, as it has to scan all channels initially. The total handoff delay is calculated using the formula, total_delay = scan_time + authentication_time + reassociation_time + load_delay + packet_delay + prediction_delay, where scan_time is the scan delay, authentication_time is the authentication delay, reassociation time is the reassociation delay, load delay is due to network load (and possible collisions/re-transmissions), packet delay is the mean delay experienced by a packet (which is a function of mean processing capacity and mean packet arrival) and prediction delay is caused by wrong next access point prediction. Overall, the graph shows a mean handoff delay of 39.2 msec for 10 paths as shown in fig. 21 and 34.5 msec for 100 mobile node paths.

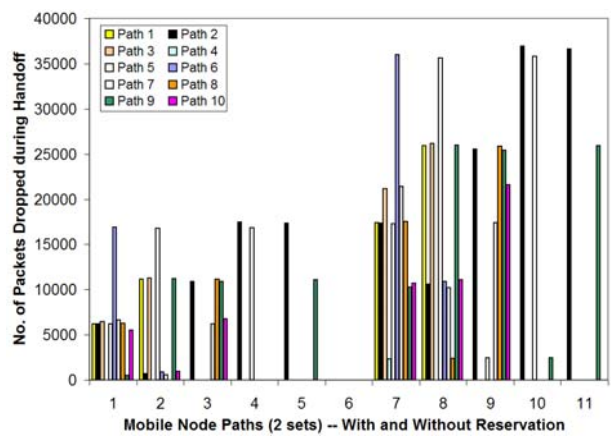

Figure 22. Bits dropped during handoff with the proposed scheme for 10 mobile node paths – with reservation (first half) and without reservation (second half)

Wireless packets dropped in handoff are a function of handoff delay and the amount of resources available. Thus the packets dropped during handoff can be reduced if the handoff delay is reduced and the available buffer size in next AP is more. As the processing capacity is 1000 packets in 1 msec, based on handoff delay above a threshold delay value, corresponding number of packets would be dropped. The data traffic dropped is shown in fig. 22, with and without reservation, where there is a great reduction in bits dropped with prior reservations done in the new approach.

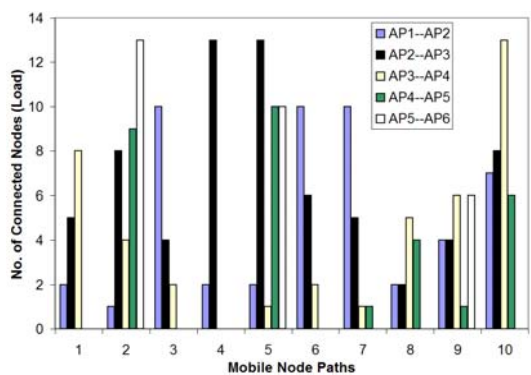

Figure 23. Load distributions in the simulation showing 10 mobile node paths

The load distribution during the mobile node movements in the ten mobile node paths considered is shown in fig. 23. The classification of the network load on a Basic Service Set (BSS) is done as follows. If there are 5 or less at an access point, the network traffic load is classified as low. If the number of attached nodes are between 5 and 10, then the network load is classified as medium. But if the number of attached nodes is equal to or greater than 10, then it is classified as heavily loaded BSS.

Fig. 24 shows the RSSI value distribution in simulation for 2 MN transitions, just before the handoff from current AP to the next AP. It can be seen that the RSSI values from the current AP is decreasing as the MN is moving away and that the RSSI values from the next AP is increasing as the MN is moving towards it. On RSSI sample 7 (around 3 to 4 mW for current AP and around 50 to 55 mW for next AP) on x-axis, the RSSI threshold warning is initiated and MN "prepares" to handoff. On RSSI sample 8 (around 1 to 2 mW for current AP and 65 to 75 mW for next AP), the RSSI handoff threshold warning is initiated and the handoff of MN happens very shortly afterwards. It should be noted that RSSI handoff threshold warning will be initiated when the current AP RSSI value falls below the RSSI handoff lower threshold value and the RSSI value of the next AP rises above the RSSI handoff higher threshold value. This is to reduce the "ping-pong" effect, if the MN moves between two access points frequently.

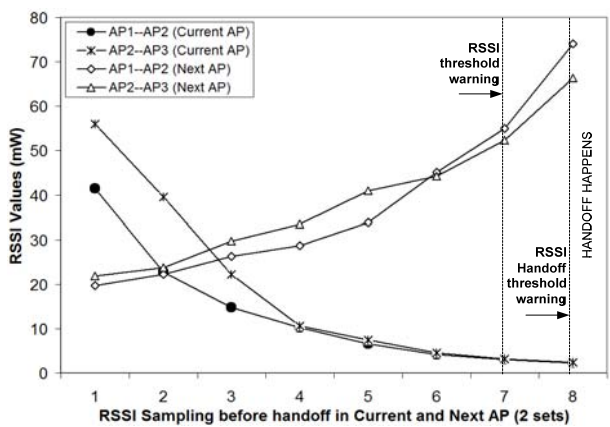

Figure 24. RSSI value distributions before MN handoff from Current AP to Next AP showing two possible handoff scenarios

## X. CONCLUSION

We have proposed a predictive mobility management scheme (PMMS) which uses mobility prediction through location tracking and data mining. The mobility prediction results suggest a prediction accuracy of 80 to 90% or more, when compared to the other baseline schemes like Transition Matrix prediction and Ignorant prediction. It also achieves fast handoff through pre-authentication, delay management and resource reservation. In simulation, the average scan delay, authentication delay and reassociation delay (using IAPP) values for 100 paths was found to be 9.9 msec, 4.3 msec and 12.2 msec respectively. The performance graphs indicate that the approach can significantly reduce the handoff delay to around 50 msec (to suit audio communication) through mobility prediction, by cutting off various delays and by using prior resource reservation.